\documentclass[12pt,preprint]{aastex}

\def\la{\mathrel{\hbox{\rlap{\hbox{\lower4pt\hbox{$\sim$}}}\hbox{$<$}}}}
\def\ga{\mathrel{\hbox{\rlap{\hbox{\lower4pt\hbox{$\sim$}}}\hbox{$>$}}}}

\slugcomment{Submitted to {\it The Astrophysical Journal (Letters)}}

\begin{document}

\title{GRB Variability-Luminosity Correlation Confirmed}

\author{Daniel E. Reichart\altaffilmark{1} and Melissa C. Nysewander\altaffilmark{1}}

\altaffiltext{1}{Department of Physics and Astronomy, University of North
Carolina at Chapel Hill, Campus Box 3255, Chapel Hill, NC 27599;
reichart@physics.unc.edu}

\begin{abstract}

Recently, Guidorzi et al.\ (2005) expanded the size of the sample of GRBs for
which variabilities and peak luminosities have been measured, from 11 to 32. 
They confirm the existence of a correlation, but find a dramatically different
relationship between $L$ and $V$ than had originally been found.  We find that
this is the result of improper statistical methodology.  When we fit a model
to the data that accommodates both statistical variance (in two dimensions)
and sample variance, we find that $L \sim V^{3.4^{+0.9}_{-0.6}}$ with a sample variance of
$\sigma_{\log{V}} = 0.20^{+0.04}_{-0.04}$, which is consistent with the original finding
of
Reichart et al.\ (2001) -- $L \sim V^{3.3^{+1.1}_{-0.9}}$ with a sample variance of
$\sigma_{\log{V}} =
0.18^{+0.07}_{-0.05}$ -- and inconsistent with the finding of Guidorzi et al.\ (2005):
 $L \propto V^{1.3^{+0.8}_{-0.4}}$ with sample variance assumed to be zero.

\end{abstract}


\section{Introduction}

A correlation between GRB variability and peak luminosity was first proposed
by Fenimore et al.\ (2000).  Motivated by
this, Reichart et al.\ (2001) constructed a robust measure of GRB variability
and computed variabilities for 18 GRBs with high time-resolution (64 msec)
light curves and variability lower limits for two GRBs with low
time-resolution (1 sec) light curves.  Of the former 18 GRBs, 11 had measured
redshifts and peak luminosities.  A correlation between $L$ and $V$ was apparent
and significant at the 3.8$\sigma$ credible level if nearby GRB 980425 was
excluded
and 4.9$\sigma$ credible level if GRB 980425 was included.  However, we did not
model
the data with a power law (as Guidorzi et al.\ 2005 claim), because the scatter
of the data around such a model is more than can be accounted for by the
data's statistical error bars alone.  This scatter is called sample variance.
 Consequently, we instead modeled the data with a normal distribution around a
line in the $\log{L} - \log{V}$ plane.  The width of this distribution, which can be
parameterized by either $\sigma_{\log{L}}$ or $\sigma_{\log{V}}$, measures the sample
variance, and is
sometimes called the ``slop'' parameter (e.g., Reichart 2001; Lee et
al.
2001; Galama et al.\ 2003; Nysewander et al.\ 2005).  Fitting such a
distribution to data with error bars in not one but two dimensions is
non-trivial.  However, we present a maximum-likelihood procedure for doing
just this in \S2.2.2 of Reichart (2001).  Applying this formalism
yielded $L \sim V^{3.3^{+1.1}_{-0.9}}$ with a sample variance of $\sigma_{\log{V}} =
0.18^{+0.07}_{-0.05}$.

Using the measure of variability introduced by Reichart et al.\ (2001),
Guidorzi et al.\ (2005) have expanded the size of the sample of GRBs for which
both variabilities and peak luminosities are known from 11 to 32, primarily
with BepposSAX GRBM data, but also with HETE-2 FREGATE, Swift BAT, CGRO BATSE,
WIND Konus, and Ulysses GRB data.  They confirm the existence of a
correlation, using Pearson's $r$, Spearman's $r_s$, and Kendall's $\tau$ tests. They
also find that the data are not well described by a power law ($\chi^2 = 1167$
or
1145, depending on how they do the fit, for only 30 degrees of freedom), but
incorrectly state that Reichart et al.\ (2001) had found this to be the case. 
Although their fits are poor, they find that $L \propto V^{1.3^{+0.8}_{-0.4}}$ or $V^{1.2^{+0.5}_{-0.2}}$,
again depending on how they do the fit, with sample variance assumed to be
zero.  This is significantly different than what Reichart et al.\ (2001) found.  

Some of the effects of not modeling sample variance when it is required can be
seen in Figures 1 and 2:  The dashed lines mark the center and 1$\sigma$
width of the
original best-fit distribution of Reichart et al.\ (2001), and appear to remain
an acceptable description of the data even though the sample has nearly
tripled in size.  We update this fit in \S2.  The dotted lines are the best-fit
power laws of Guidorzi et al.\ (2005), which do not appear to acceptably
describe even the trend of the data.  

\section{Updated Fits}

We now refit the model of Reichart et al.\ (2001) to the expanded data set of
Guidorzi et al.\ (2005).  However, we do not include GRB 000210 in this fit: 
This GRB is missing high time-resolution data for 2.5 sec during the peak of
this $T_{90} = 8.1$ sec long event and consequently should be viewed as only
providing a lower limit to the variability.  Indeed, this GRB is significantly
less variable than all other GRBs of comparable peak luminosity, typically by
an order of magnitude.  Using the statistical formalism presented in Reichart
(2001) for fitting distributions to data with error bars in two dimensions, we
find that $L \sim V^{3.4^{+0.9}_{-0.6}}$ with a sample variance of
$\sigma_{\log{V}} = 0.20^{+0.04}_{-0.04}$ 
if we exclude GRB 980425 from the fit and $L \sim V^{3.5^{+0.9}_{-0.6}}$ with a sample
variance of $\sigma_{\log{L}} = 0.20^{+0.04}_{-0.04}$ if we include GRB 980425.  The solid
lines
in Figures 1 and 2 mark the centers and 1$\sigma$ widths of the best-fit
distributions, respectively.  These distributions are in excellent agreement
with out earlier work.

If we also include GRB 000210, we find that $L \sim V^{4.0^{+1.5}_{-0.9}}$ with a sample
variance of $\sigma_{\log{L}} = 0.26^{+0.05}_{-0.04}$.  This shows that the difference
between
our fits and those of Guidorzi et al.\ (2005) are not due to our exclusion of
this point, but due to their exclusion of sample variance in their model.  

\section{Conclusions}

Using a statistical formalism that is appropriate for this data set, we fit a
distribution in the  $\log{L} - \log{V}$ plane to the expanded data set of Guidorzi
et al.\ (2005).  Our findings are in excellent agreement with our earlier work,
when the sample was approximately one-third its current size.  The
variability-luminosity correlation is now best described by $L \sim V^{3.4^{+0.9}_{-0.6}}$ with a sample variance of 
$\sigma_{\log{V}} = 0.20^{+0.04}_{-0.04}$.

\acknowledgements
DER thanks Don Lamb for useful discussions and very gratefully acknowledges
support from NSF's MRI, CAREER, PREST, and REU programs, NASA's APRA, Swift
GI
and IDEAS programs, and especially Leonard Goodman and Henry Cox.

\clearpage

\figcaption[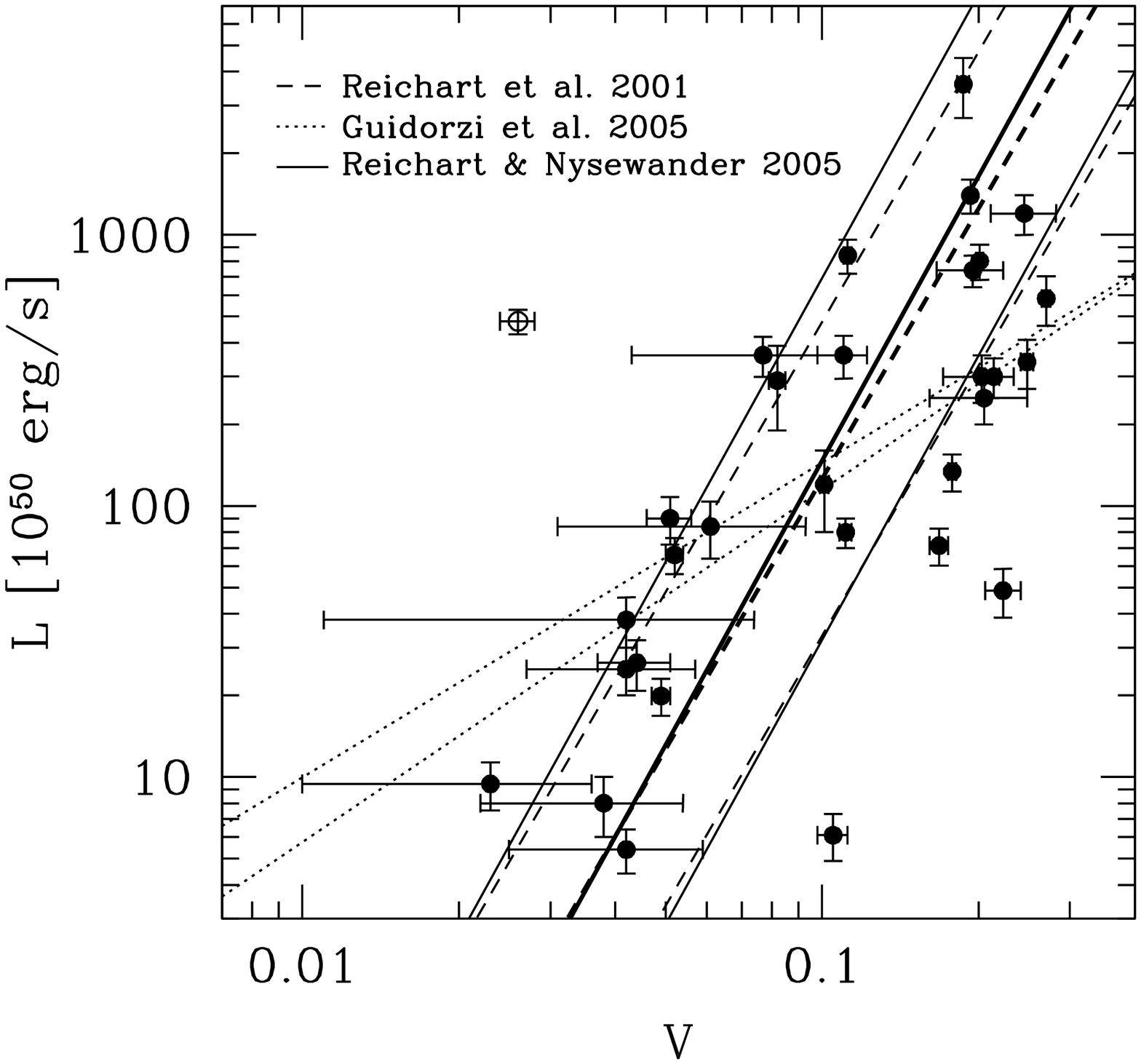]{Variabilities $V$ and peak luminosities $L$ of all of the GRBs in the
sample of Guidorzi et al.\ (2005) except for nearby GRB 980425.  Dashed lines
mark the center and 1$\sigma$ width of the original best-fit
distribution of Reichart
et al.\ (2001) and solid lines mark the center and 1$\sigma$ width of the
updated
best-fit distribution of this paper.  Dotted lines are the best-fit power laws
of Guidorzi et al.\ (2005).  The variability of the unfilled circle (GRB
000210) should be treated as a lower limit and consequently we do not include
it in our updated fit, though we show that including it does not significantly
change our results (\S2).}

\figcaption[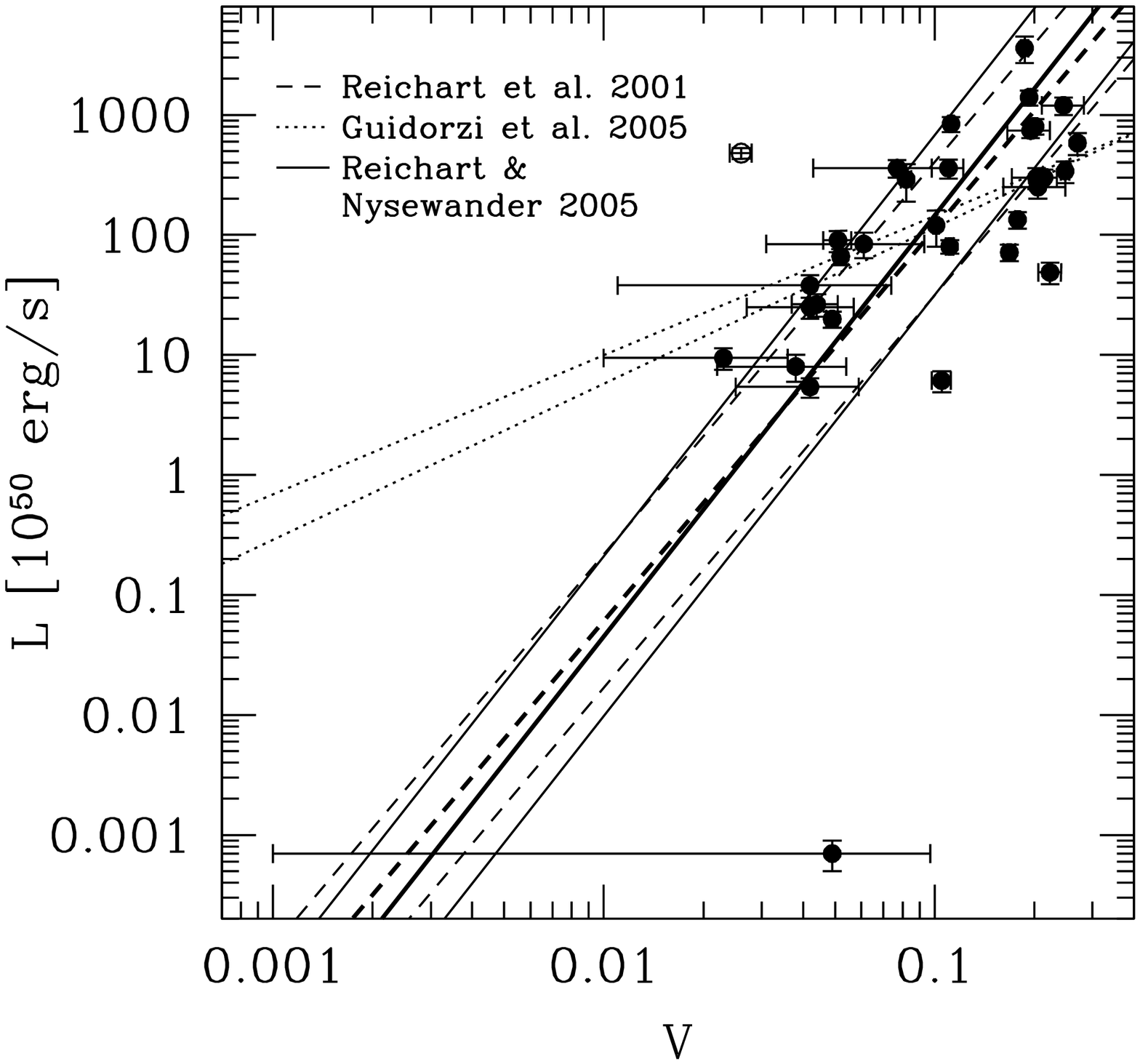]{Same as Figure 1, except including GRB 980425.}

\clearpage

\setcounter{figure}{0}

\begin{figure}[tb]
\plotone{f1.eps}
\end{figure}

\begin{figure}[tb]
\plotone{f2.eps}
\end{figure}


\clearpage

\begin{thebibliography}{}

\bibitem[Fenimore \& Ramirez-Ruiz(2000)]{fr00} Fenimore, E.~E., 
\& Ramirez-Ruiz, E. 2000, preprint (astro-ph/0004176)

\bibitem[Galama et al.(2003)]{2003ApJ...587..135G} Galama, T.~J., et al.\ 
2003, \apj, 587, 135

\bibitem[Guidorzi et al.(2005)]{gfm+05} Guidorzi, C., et al.\ 2005,
preprint (astro-ph/0507588)

\bibitem[Lee et al.(2001)]{2001ApJ...561..183L} Lee, B.~C., et al.\ 2001, 
\apj, 561, 183 

\bibitem[Nysewander et al.(2005)]{nrp+05} Nysewander, M.~C., et al.\ 2005,
preprint (astro-ph/0505474)

\bibitem[Reichart(2001)]{2001ApJ...553..235R} Reichart, D.~E.\ 2001, \apj,
553, 235

\bibitem[Reichart et al.(2001)]{2001ApJ...552...57R} Reichart, D.~E., et al.\
2001, \apj, 552, 57

\end{thebibliography}
\end{document}